\documentstyle[12pt]{article}

\bigskip
\title{ Weyl Pair, Current Algebra and Shift Operator }
\author{Zhan-Ning Hu \thanks{\bf email address: huzn@itp.ac.cn}
\thanks{This research was partially supported by the China center of advanced
science and technology.}
\\ Institute of Theoretical Physics, Academia Sinica \\
  P. O. Box 2735, Beijing 100080, China \thanks{\bf mail address}}
\date{Oct. 3, 1994}
\begin{document}
\maketitle
\bigskip

\begin{abstract}
The Abelian current algebra on the lattice is given from a series of  the
independent Weyl pairs and the shift operator is constructed by this algebra.
So the realization  of the operators of the braid group is obtained. For
$|q|\neq 1$ the shift operator is the product of the theta functions of the
generators $w_n$ of the current algebra. For $|q|=1$ it can be expressed by the
quantum dilogarithm of $w_n$.
\smallskip

\bigskip

\end{abstract}
\newpage

\section{Introduction}
 Owing to the discrete space-time picture much benefit to the lattice
sine-Gordon model ones try to discretize the conformal filed theory where the
lattice Virasoro algebra plays an important role \cite{Ba,FV,Ge}.  One of the
approaches is given by Faddeev and Volkov \cite{FV}. They start with the
periodic free field (Abelian current) which can be gotten from the continuous
limit of the following current algebra:
$$
w_{n-1}w_n=q^2w_nw_{n-1},~~~n=2,3,\cdots,2N,
$$
\begin{equation} {\label 1}
w_{2N}w_1=q^2w_1w_{2N},
\end{equation}
$$
w_mw_n=w_nw_m, 1<|m-n|<2N-1,
$$
for the operators $w_n (n=1,2,\cdots,2N)$ where $q$ is a complex constant. The
key step in this approach is to construct the shift operator and Ref. [2] has
discussed it with $|q|<1$. What is about the case of $|q|\geq 1$ is just the
aim of this letter.

In section 2 the $w$ algebra mentioned above is denoted by a set of independent
Weyl pairs and the shift operator is given for $|q|<1$. The different initial
values of the products in the shift operator mean some factors times $w_n$. The
case for $|q|>1$ is discussed in section 3. We construct the shift operator by
the inverse of it. In the above two cases the shift operator can be always
denoted in the form of the theta functions \cite{book}. In section 4 we
constructed the shift operator by using the quantum dilogarithm \cite{fk,Ki}
which appeared in two and three-dimensional lattice models \cite{b2,k12,h2}.

\section{Weyl Pair and Shift Operator with $|q|<1$}
 From $N-1$ independent Weyl pairs $x_i, y_i (i=1,2,\cdots,N-1)$ with the
relations
\begin{equation}
\begin{array}{l}
x_iy_j=q^{2\delta_{ij}}y_jx_i,~~x_ix_j=x_jx_i,\\
y_iy_j=y_jy_i,~~~1\leq i,j\leq N-1,
\end{array}
\end{equation}
where $\delta_{ij}$ is the Kronecker symbol and $q$ is a complex constant, we
can get the current algebra (1) by setting
\begin{equation}
\begin{array}{l}
w_{2i+1}=x^{-1}_ix_{i+1},~~i=0,1,2,\cdots,N-1, \\
w_{2j}=y_j,~~j=1,2,\cdots,N-1, \\
w_{2N}=\prod^{N-1}_{l=1}y^{-1}_l.
\end{array}
\end{equation}
Let
\begin{equation} {\label c}
C_1=\prod w_1w_3\cdots w_{2N-1},~~ C_2=\prod w_2w_4\cdots w_{2N}.
\end{equation}
They are the two central elements of the $w$ algebra. If $C_1\neq C_2$, by
making the transformations of
\begin{equation}
w'_1=C_2w_1,~~w'_{2N}=C_1w_{2N},~~w'_i=w_i,~~ 1<i<2N,
\end{equation}
we can get the $w'$ algebra with the equal central elements $C'_1=C'_2=C_1C_2$.
So we always assume that the two central elements of the $w$ algebra are equal,
$i.e. C_1=C_2$.

Now we construct the shift operator with $|q|<1$. Letting
\begin{equation} \label{2.1}
h_n=\psi_a(w_n,\alpha)\psi_b(w^{-1}_n,\beta), ~~n=1,2,\cdots,2N,
\end{equation}
where
\begin{equation}
\psi_a(w_n,\alpha)=\prod^{\infty}_{j=a}(1+w_n\alpha
q^{2j-1}),~~\psi_b(w^{-1}_n,\beta)=\prod^{\infty}_{j=b}(1+w^{-1}_n\beta
q^{2j-1}),
\end{equation}
we get that
\begin{equation} {\label w}
w_nh_{n-1}h_n=h_{n-1}h_nw_{n-1},~~n=2,3,\cdots,N,
\end{equation}
under the restrict condition of
\begin{equation} {\label r}
\alpha \beta q^{2(a+b-2)}=1
\end{equation}
for the integers $a,b$ and the parameters $\alpha, \beta$. Then by setting
\begin{equation} {\label u}
U=h_1h_2\cdots h_{2N-1}
\end{equation}
we have that
\begin{equation}
w_nU=Uw_{n-1}
\end{equation}
for $n=1,2,\cdots,2N$ with $w_0=w_{2N}$ where $C_1=C_2$ has been used. So $U$
is a shift operator. By considering the relation (\ref{r}) we have
\begin{equation}
h_n=G^{-1}\vartheta_3(w_n\alpha q^{2(a-1)})
\end{equation}
where \cite {book}
\begin{equation} \label{theta}
\vartheta_3(v)=G\prod^{\infty}_{n=1}(1+q^{2n-1}v)\prod^{\infty}_{n=1}(1+q^{2n-1}v^{-1}),~~G=\prod^{\infty}_{n=1}(1-q^{2n}).
\end{equation}
{}From Eq. (\ref{w}) we have
\begin{equation}
h_nh_{n-1}h_n=h_{n-1}h_nh_{n-1},~~n=1,2,\cdots,2N
\end{equation}
where $h_0=h_{2N}$. This gives the representation of the operators of the braid
group. By making the notation of
\begin{equation} {\label A}
\psi(A)=\prod^{\infty}_{j=1}(1+Aq^{2j-1}),
\end{equation}
$\psi_a(x,\alpha)$ and $\psi_b(
y,\beta)$ have the property
\begin{equation} {\label m}
\psi_b(y,\beta)\psi_a(x,\alpha)=\psi(x\alpha q^{2(a-1)}+y \beta q^{2(b-1)}+qyx)
\end{equation}
for the Weyl pair $x,y$ with the relation $xy=q^2yx$. Then the shift operator
can be expressed as
$$
U=\psi (w^{-1}_1\beta q^{2(b-1)}+w_2\alpha q^{2(a-1)}+qw^{-1}_1w_2)\psi
(w^{-1}_2\beta q^{2(b-1)}+w_3\alpha q^{2(a-1)}+qw^{-1}_2w_3)
$$$$
{}~~~~~~~~~~\cdots \psi (w^{-1}_{2N-1}\beta q^{2(b-1)}+w_{2N}\alpha
q^{2(a-1)}+qw^{-1}_{2N-1}w_{2N})
$$$$
{}~~=\psi (w_1\alpha q^{2(a-1)}+w^{-1}_2\beta q^{2(b-1)}+qw_1w^{-1}_2)\psi
(w_2\alpha q^{2(a-1)}+w^{-1}_3\beta q^{2(b-1)}+qw_2w^{-1}_3)
$$
\begin{equation} \label{psi}
{}~~~~~~~~~~\cdots \psi (w_{2N-1}\alpha q^{2(a-1)}+w^{-1}_{2N}\beta
q^{2(b-1)}+qw_{2N-1}w^{-1}_{2N})
\end{equation}
It is just the expression given by Faddeev and Volkov. This result  means that
the initial values $a,b,\alpha$ and $\beta$ with the relation (\ref{r}) denote
the factor $\alpha q^{2(a-1)}$ times $w_n$ in the shift operator. And the
operator $U$ constructed by Eq. (\ref u) with the relation (\ref{2.1}) is the
product of the theta functions of the $w_n$.
Similarly as the Eq. (\ref{m}), using the mathematical reduction we can prove
further that
\begin{equation} \label {222}
\prod^n_{j=1}(\beta+yq^{2j-1})\prod^n_{j=1}(\alpha+xq^{2j-1})=\prod^n_{j=1}(\alpha \beta+Yq^{2j-1}),~~n\geq 1,
\end{equation}
for the arbitrary $\alpha, \beta$ and $q$ where $Y=\beta x+\alpha y+qyx$.

\section{The Shift Operator with $|q|>1$}

Setting
\begin{equation}
h_n=\bar{\psi}_a(w_n,\alpha)\bar{\psi}_b(w^{-1}_n,\beta)
\end{equation}
where
\begin{equation}
\bar{\psi}_a(w_n,\alpha)=\prod^{\infty}_{j=a}(1+w_n\alpha \bar q ^{2j-1}),~~
\bar{\psi}_b(w^{-1}_n,\beta)=\prod^{\infty}_{j=b}(1+w^{-1}_n\beta \bar q
^{2j-1})
\end{equation}
with the notation $\bar q =1/q$, from Eq. (1), we have
\begin{equation} \label{hh}
w_nh_{n+1}h_n=h_{n+1}h_nw_{n+1},~~n=1,2,\cdots,2N-1,
\end{equation}
when the integers $a,b$ and the parameters $\alpha,\beta$ satisfy the relation:
$\alpha\beta\bar q ^{2(a+b-2)}=1$. In this way, $h_n$ is given also by the
theta function
\begin{equation}
h_n=G^{-1}\vartheta_3(w_n\alpha \bar q ^{2(a-1)})
\end{equation}
where $\vartheta_3$ is defined by relation (\ref{theta}) with the substitution
of $\bar q$ for $q$. And the representation of the operators of the braid group
is also given. Set
\begin{equation} \label{u-}
U^{-1}=h_{2N}h_{2N-1}\cdots h_2
\end{equation}
By considering that  the two central elements $C_1,C_2$, of the current algebra
(1), are equal, from Eq. (\ref{hh}), we have
\begin{equation}
w_nU^{-1}=U^{-1}w_{n+1},~~n=1,2,\cdots,2N
\end{equation}
with $w_{2N+1}\equiv w_1$. So operator $U$ expressed by Eq. (\ref{u-}) is
indeed the shift operator for $|q|>1$. Furthermore, it can be expressed as
$$
U^{-1}=\bar{\psi}(w^{-1}_{2N}\beta q^{2(1-b)}+w_{2N-1}\alpha
q^{2(1-a)}+qw_{2N-1}w^{-1}_{2N}) ~~~~~~~~~~~~~~~
$$$$
\bar{\psi}(w^{-1}_{2N-1}\beta q^{2(1-b)}+w_{2N-2}\alpha q^{2(1-a)}+qw_{2N-2}w^
{-1}_{2N-1})
$$$$
\cdots\bar{\psi}(w^{-1}_{2}\beta q^{2(1-b)}+w_{1}\alpha
q^{2(1-a)}+qw_{1}w^{-1}_{2})
$$$$
=\bar{\psi}(w_{2N}\alpha q^{2(1-a)}+w^{-1}_{2N-1}\beta
q^{2(1-b)}+qw^{-1}_{2N-1}w_{2N}) ~~~~~~~~
$$$$
\bar{\psi}(w_{2N-1}\alpha q^{2(1-a)}+w^{-1}_{2N-2}\beta
q^{2(1-b)}+qw^{-1}_{2N-2}w_{2N-1})
$$
\begin{equation}
\cdots\bar{\psi}(w_{2}\alpha q^{2(1-a)}+w^{-1}_{1}\beta
q^{2(1-b)}+qw^{-1}_{1}w_{2})
\end{equation}
where $\bar{\psi}$ is defined as
$\bar{\psi}(A)=\prod^{\infty}_{j=1}(1+Aq^{1-2j})$ for operator $A$. Then we get
the shift operator with $|q|>1$ by considering the inverse of it. It should be
noted that  the above relation can be obtained also by defining that
$U^{-1}=h_{2N-1}h_{2N-2}\cdots h_1$.

\section{The Shift Operator with $|q|=1$}

Set
\begin{equation}
q^2=\omega=exp(2\pi i/L), ~~\omega^{1/2}=exp(\pi i/L).
\end{equation}
{}From Eq. (\ref{222}) we get that
\begin{equation} \label{sum}
(x+y)^N=x^N+y^N,~~xy=\omega yx,
\end{equation}
by considering the relation \cite{phD,h1}
\begin{equation}
\prod^L_{j=1}(1-x\omega^j )=1-x^N.
\end{equation}
Now we fix the $L$-th powers of the generators $w_n$ of the current algebra (1)
are the identity operator:
\begin{equation} \label{cho}
w^L_n=1,~~n=1,2,\cdots,2N,
\end{equation}
since they are the central elements of the $w$ algebra in this case.
Introducing the operators
\begin{equation}
W^{(1)}_n=k(w^{-1}_n+w_{n+1}-\omega^{1/2}w^{-1}_nw_{n+1}),~~
W^{(2)}_n=k(w_n+w^{-1}_{n+1}-\omega^{1/2}w_nw^{-1}_{n+1}),
\end{equation}
where $n=1,2,\cdots,2N-1$ and $k^L=1/3$ we have that
\begin{equation}
{W^{(1)}_n}^L={W^{(2)}_n}^L=1
\end{equation}
by taking account of the relations (\ref{sum}) and (\ref{cho}). The spectrum of
any operator, $A$, whose $L$-th power is the identity operator, is given by $L$
distinct numbers
\begin{equation} \label{spe}
\omega^l,~~l=0,1,\cdots,L-1.
\end{equation}
Then all of the operators $w_n, W^{(1)}_n,W^{(2)}_n, (n=1,2,\cdots,2N-1)$ and
$w_{2N}$ have one and the same spectrum (\ref{spe}). Define \cite{k12,h2}
\begin{equation}
w(a,b,c|l)=\prod^l_{j=1}b/(c-a\omega^j),~~a^L+b^L=c^L,~~n\geq 0,
\end{equation}
with $w(a,b,c|0)=1$. The quantum dilogarithm $\Psi(A)$ \cite{fk} depending on
the operator $A$ which has the spectrum (\ref{spe}) can be defined as an
operator commuting with $A$ and has the spectrum
\begin{equation}
\Psi(\omega^l)=\Psi(1)w(a,b,c|l)
\end{equation}
where $\Psi(1)$ is a non-zero complex factor. So the "functional" relation of
the quantum dilogarithm $\Psi(A)$ has the following form:
\begin{equation}
\Psi(\omega^{-1}A)\Psi(A)^{-1}=(c-aA)/b.
\end{equation}
It determines the operator $\Psi(A)$ up to a complex factor. By setting
\begin{equation}
c^2=\omega a^2
\end{equation}
it can be proved easily that the operator $U$ defined by Eq. (\ref u) with
\begin{equation}
h_n=\Psi(w_n)\Psi(w^{-1}_n),~~n=1,2,\cdots,2N-1,
\end{equation}
 is the shift operator for $|q|=1$. And $h_n$ gives also the representation of
the braid group. If we make the limit of:
$$
L\rightarrow\infty,~~b=2^{1/L}\rightarrow 1,~~k=3^{-1/L}\rightarrow 1,
$$
the "functional" relation of the quantum dilogarithm $\Psi(A)$ can be written
as
\begin{equation}
\Psi(\omega^{-1}A)\Psi(A)=1-\omega^{-1/2}A
\end{equation}
where
$$
A=w_n,~~w^{-1}_n+w_{n+1}-\omega^{1/2}w^{-1}_nw_{n+1},~~w_n+w^{-1}_{n+1}-\omega^{1/2}w_nw^{-1}_{n+1},~~w_{2N},
$$
with $n=1,2,\cdots,2N-1$. Then in this limiting the shift operator $U$ for
$|q|=1$ can be expressed also in the form (\ref{psi}) with  substituting $\psi
(w^{-1}_n\beta q^{2(b-1)}+w_{n+1}\alpha q^{2(a-1)}+qw^{-1}_nw_{n+1})$ and $\psi
(w_n\alpha q^{2(a-1)}+w^{-1}_{n+1}\beta q^{2(b-1)}+qw_nw^{-1}_{n+1})$ by $ \Psi
(w^{-1}_n+w_{n+1}+\omega^{1/2}w^{-1}_nw_{n+1})$ and $\Psi
(w_n+w^{-1}_{n+1}+\omega^{1/2}w_nw^{-1}_{n+1}), n=1,2,\cdots,2N-1,$
respectively.

As the conclusion, we get the shift operator with $q$ on a complex plane from
the current algebra (1) which can be constructed from a series of independent
Weyl pairs and can be reduced to the periodic free field in a continues limit.
So the representation of the operators of the braid group is obtained. The
integers $a,b$ and the parameters $\alpha,\beta$ in $h_n$ mean some factors
appearing for $w_n$ when it denotes the shift operator which is the product of
the theta functions of the operators $w_n$ with $|q|\neq 1$. When $|q|>1$ we
get the shift operator by considering the inverse of it. The shift operator
with $|q|=1$ is expressed by the quantum dilogarithm \cite{Ki} which appeared
also in Baxter-Bazhanov model \cite{fk,b2}. So it is a interesting subject to
discuss the lattice Virasoro algebra and construct the shift operator from the
respect of the chiral Potts model \cite{h2}.

\section*{Acknowledgment}
The author would like to thank H. Y. Guo and B. Y. Hou  for the interesting
discussions.


\newpage

\begin{thebibliography}{11}

\bibitem{Ba} O. Babelon, Phys. Lett. B238 (1990) 234.

\bibitem{FV} L. Faddeev and A. Yu. Volkov, Abelian current algebra and the
Virasoro algebra on the lattice, HU-TFT-93-29* (1993).

\bibitem{Ge} J. -L. Gervais, Phys. Lett., B160 (1985) 277; B160 (1985) 279.

\bibitem{book} K. Chandrasekharan, Elliptic function, (Grundlehren der
mathematischen Wissenschaften 281), (Germany, 1985).

\bibitem{fk} L. D. Faddeev and R. M. Kashaev, Quantum Dilogarithm, Preprint,
September (1993).

\bibitem{Ki} A. N. Kirillov, Dilogarithm Identities, Preprint, hep-th/9408113.

\bibitem{b2} V. V. Bazhanov and R. J. Baxter, J. Stat. Phys., 71 (1993) 839.

\bibitem{k12} R. M. Kashaev, V. V. Mangazeev and Yu. G. Stroganov, Int. J. Mod.
Phys., A8 (1993) 587; A8 (1993) 1399.

\bibitem{h2} Z. N. Hu, Mod. Phys. Lett., B8 (1994) 779.

\bibitem{phD} Z. N. Hu, Exactly Solved Models in Statistical Mechanics and
Bosonization of $su(3)$ Parafermion, Ph. D Thesis, February (1994).

\bibitem{h1} Z. N. Hu, Three-Dimensional Star-Star Relation, to appear in Int.
J. Mod. Phys. A.


\end{thebibliography}
\end{document}